\documentclass{easychair}

\usepackage{doc}

\usepackage{subfigure}
\usepackage{array}
\usepackage{sidecap}

\newcolumntype{V}{>{$\vcenter\bgroup\hbox\bgroup}c<{\egroup\egroup$}}
\def\Hline{\noalign{\hrule height 4\arrayrulewidth}}

\title{Colorectal Cancer Outcome Prediction from H\&E Whole Slide Images using Machine Learning and Automatically Inferred Phenotype Profiles}

\author{Xingzhi Yue
\and Neofytos Dimitriou
\and Peter D.\ Caie \and David J.\ Harrison \and Ognjen Arandjelovi\'c
}

\institute{School of Computer Science \& School of Medicine\\University of St Andrews, UK}

\authorrunning{Yue, Dimitriou, and Arandjelovi\'c}

\titlerunning{Colorectal Cancer Outcome Prediction from H\&E Whole Slide Images}

\begin{document}

\maketitle

\begin{abstract}
  Digital pathology (DP) is a new research area which falls under the broad umbrella of health informatics. Owing to its potential for major public health impact, in recent years DP has been attracting much research attention. Nevertheless, a wide breadth of significant conceptual and technical challenges remain, few of them greater than those encountered in the field of oncology. The automatic analysis of digital pathology slides of cancerous tissues is particularly problematic due to the inherent heterogeneity of the disease, extremely large images, amongst numerous others. In this paper we introduce a novel machine learning based framework for the prediction of colorectal cancer outcome from whole digitized haematoxylin \& eosin (H\&E) stained histopathology slides. Using a real-world data set we demonstrate the effectiveness of the method and present a detailed analysis of its different elements which corroborate its ability to extract and learn salient, discriminative, and clinically meaningful content.
\end{abstract}

\setcounter{tocdepth}{2}

\vspace{-0.3pt}\section{Introduction}\vspace{-0.3pt}
Colorectal cancer (CRC), also known as bowel cancer or colon cancer, is the development of cancer from the colon or rectum and accounts for about 10\% of all cancer cases worldwide \cite{Kuip2015}. Precise diagnosis and prognosis are very important in the choice of the most appropriate treatment and the facilitation of the subsequent clinical management of patients. Consequently, an increasing amount of current research is concerned with improving nuanced disease understanding, and the precision and accuracy of survivability estimation.

The Tumour-Node-Metastasis (TNM) staging system is still widely regarded as one of the best population level predictors of CRC outcome \cite{fleming2012colorectal}. However, its patient level precision is low \cite{pages2018international}. In part driven by this observation and the desire to effect improvement, the past five years have witnessed a major growth of the application of various machine learning techniques in patient level cancer prognosis \cite{chen2014risk,TunAranCaie2018,NeoAranHarrCaie2018}. 

The aim of the present work is to develop an end-to-end framework that predicts the disease specific survivability of CRC patients over a five year period by applying deep learning to unannotated whole slide histopathology images (WSIs). The key challenges addressed are thus as follows:
\begin{itemize}
  \item Computational impracticality of training classifiers with gigapixel WSIs,\\[-15pt]
  \item Efficient ground-truth labelling of individual image patches,\\[-15pt]
  \item Determination of discriminative patch subsets,\\[-15pt]
  \item Convolutional neural network (CNN) design for patch level outcome prediction, and\\[-15pt]
  \item Fusion of patch level into image level predictions.
\end{itemize}

\section{Relevant background}\vspace{-0.3pt}
In 2018 in the United States, the estimated number of new cases of CRC was 97,220 with 50,630 disease associated deaths \cite{ACS2018}. The overall death rate of CRC has decreased from 28 per 100,000 (1975) to 14 per 100,000 (2015) as the result of increased screening, decline in incidence, and improvement of treatment \cite{ACS2018}. Screening can effectively prevent CRC by the identification and subsequent removal of early-stage precancerous growth, with a range of factors employed to identify patients at risk. Some of the notable risk factors include colorectal adenomas \cite{johns2001systematic}, hereditary conditions including Lynch syndrome and adenomatous polyposis \cite{mork2015high}, personal history of long-standing chronic ulcerative colitis \cite{laukoetter2011intestinal}, and alcohol use \cite{fedirko2011alcohol}. Although colorectal cancer screening guidelines do not distinguish between female and male, the statistics show that the number of new cases of males is 17\% greater than that of female \cite{liang2009cigarette} while female over 65 years old presents higher mortality rate and lower 5-year survival rate of CRC compared to their age-matched male counterparts \cite{kim2015sex}.

However, CRC is found to be highly treatable and often curable when it is confined to the bowel after surgical intervention \cite{miller2016cancer}. Nevertheless, CRC is generally considered not curable when the cancer cells have spread to other organs, termed metastasis. In this case, appropriate choices of health management, including chemotherapy or targeted therapy, can still help improve the quality and length of life \cite{miller2016cancer}. Therefore, the early diagnosis and accurate prognostic prediction of the cancer aggressiveness and patient outcome are significant.

\vspace{-0.3pt}\section{CRC staging and prognosis}\vspace{-0.3pt}
In present clinical practice, the main prognostic factors for CRC comprise: (T) depth of tumour penetration through bowel wall, (N) presence or absence of nodal involvement, and (M) presence or absence of distant metastases. These form the basis of the five stage TNM staging system \cite{gospodarowicz2017tnm}.  Stage 0 is least severe, with all the lesions restricted to the mucosa and the lamina propria. Local excision or simple polypectomy with clear margins is the most common treatment option. In Stage~I, cancer may have grown into the muscularis mucosa or into the muscularis propria but has not spread deeper into the colon muscle wall, to nearby lymph nodes or other distant sites. Because CRC at this stage is still localized, it also has a high cure rate with wide surgical resection and anastomosis. Stage~II characterizes CRC that has spread to to or beyond the serosa and may have grown into nearby tissue or organs, but not to the lymph nodes and has not metastasised.  Surgical resection is again the standard treatment, however high-risk patients, such as those with t4 disease may be offered chemotherapy. Stage III is characterized by lymph node involvement and the standard treatments are wide surgical resection and anastomosis, and adjuvant chemotherapy. Stage IV disease is characterized bymetastatic disease. The treatment of CRC at this stage largely depends on the sites of metastatic disease. Liver metastasis makes up approximately 50\% of Stage IV and recurrent CRC, and the options for treatment include all the aforementioned ones as well as palliative radiation therapy, palliative chemotherapy, and targeted therapy. 

\subsection{H\&E staining}
\begin{figure}[t]
  \centering 
  \includegraphics[clip,trim={15pt 2pt 0 0},width=.45\textwidth]{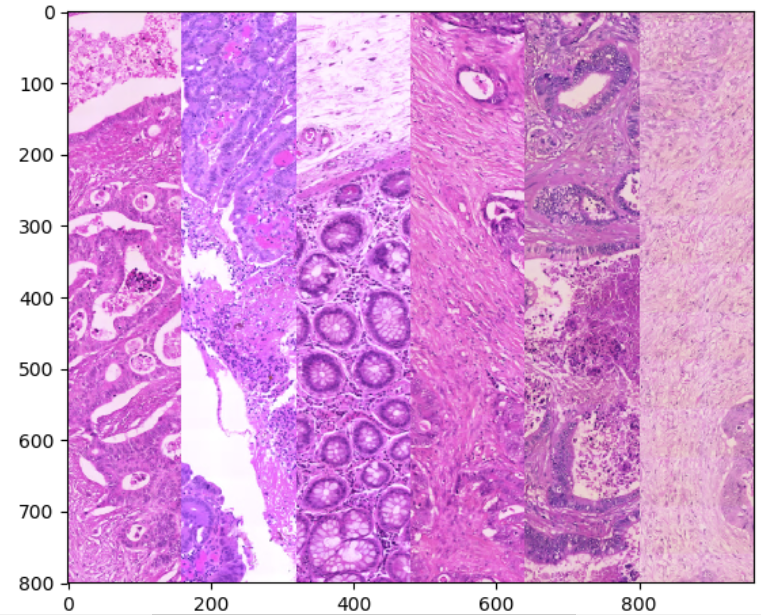}
  \includegraphics[width=.45\textwidth]{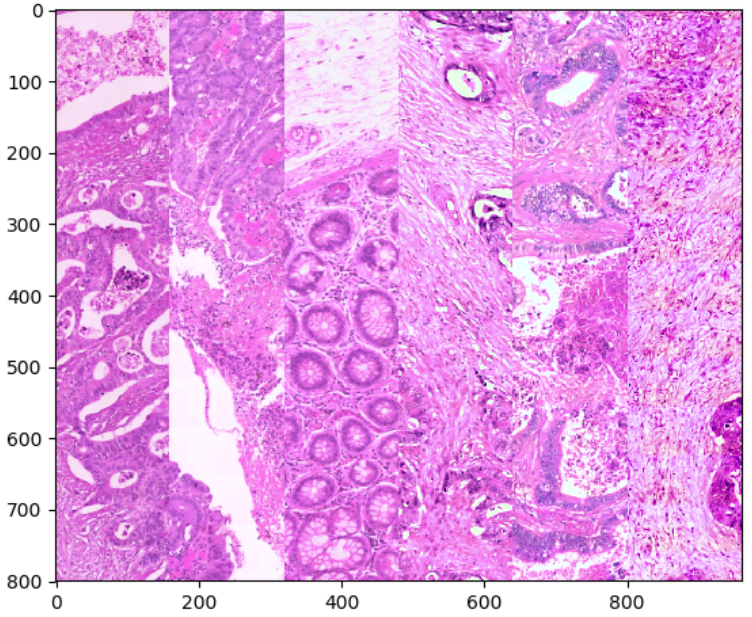}
  \caption{Chromatic normalization examples (left \& right: original \& normalized tiled strips).}
  \label{f:normEx}
\vspace{-0.10pt}\end{figure}

In this work, we utilize digitized whole slide images of archived diagnostic histopathological tissue sections stained with haematoxylin and eosin (H\&E); see examples in Fig~\ref{f:normEx}. Eosin is an acidic dye that stains the basic structures red or pink, such as the proteins within the cytoplasm. Haematoxylin, on the other hand, is a basic dye which stains the acidic structure blue or purple, such as DNA in the nucleus.

Within an H\&E stained colon section, we can observe nuclei of cells in purplish blue, cytoplasm in red, erythrocytes in cherry red, collagen and mitochondria in pale pink. However, the colour intensity of the stain depends on both the amount of stain applied and the duration of exposure.

\vspace{-0.5pt}\subsection{Previous work}\vspace{-0.3pt}
The manual reporting of H\&E stained tissue sections under the microscope, or WSIs on the computer monitor, for TNM staging is laborious. It is also largely based on the subjective experience-based assessment of the pathologist, which often causes variation across different observers. Motivated by these limitations, there is an increasing amount of research on the use of machine learning for the analysis of WSIs.

As a data-driven and end-to-end approach that learns high-level feature without subjective biases, application of convolutional neural networks (CNNs), computer-aided interpretation tools, can be traced to 1990s when a convolutional artificial neural network which attempts to mimic a radiologist's analytic process of radiographic images was introduced \cite{lo1995artificial}. Thereafter, CNN architectures for carotid intima-media thickness measurement in ultrasound images \cite{shin2016automating}, brain tumour segmentation in magnetic resonance imaging scans \cite{havaei2017brain}, neuronal membrane segmentation in electron microscopy images \cite{ciresan2012deep}, and many others have been proposed. For WSI histological analysis, CNNs have been used for challenging problems such as automated nuclear atypia scoring \cite{lu2017deep} and the discrimination between epithelial and stromal tissues \cite{bianconi2015discrimination}, with promising results.

More recently, a series of works suggested that fine-tuned pre-trained CNNs outperformed CNNs that are trained from scratch while taking much less time as well \cite{gao2018holistic}. Conventionally, when training from scratch, all the parameters in the architecture of an artificial neural network are randomly initialized. By contrast, when fine-tuning a CNN, the weight and bias values are initialized with the parameters of a pre-trained CNN with the same architecture. It is argued that the early layers of CNN learn the low-level image features which are roughly the same for different vision problems, however, the deeper layers that learn the high-level features are specific with respect to the classification task itself \cite{tajbakhsh2016convolutional}. Four specific image analysis problems, including colonic polyp or pulmonary embolism detection, colonoscopy frame classification and intima-media boundary segmentation were considered by Tajbakhsh et al.\ \cite{tajbakhsh2016convolutional} to demonstrate the potential of knowledge transfer between `natural' and medical images.

\subsubsection{Patch-based CNNs}
Although CNNs are widely considered as the state-of-the-art models in various applications of image classification, the analysis of WSIs remains challenging because training a deep CNN model with gigapixel WSIs is still computationally impractical. Hence, most of the aforementioned techniques work with severely down-sampled images. However, this approach inherently effects a loss of discriminative information at finer scales.

Hou et al. \cite{hou2016patch} train a model on patches of high-resolution images and from these make predictions for entire WSIs, with an expectation maximization based algorithm used to automatically determine discriminative patches. Thereafter, a number of other methods which rely on the use of image patches and CNNs have demonstrated promising results in distinguishing WSIs of tumorous and normal tissues \cite{jamaluddin2017tumor}, as well as the segmentation of precursor lesions \cite{albayrak2017segmentation}. However, prognosis is an inherently more difficult learning challenge \cite{mackillop2003importance}.

Recently an adaptive sampling method was applied in an end-to-end framework \cite{zhu2017wsisa} to cluster group of images with different local content. This framework comprises four key stages: (i) adaptive generation of patches from WSIs, (ii) patch clustering according to the phenotypes, (iii)
automatic clustering selection, and (iv) aggregation of cluster level predictions. However, due to the low-resolution of down-sampled phenotypes and the lack of an effective way to aggregate patch-wise predictions, the framework only achieves 57\% average accuracy. The framework we propose in the present paper overcomes the key limitations of the aforementioned work in part by introducing a novel way of generating phenotypes.

\section{Proposed methodology}
The main objective of this work is to predict the survivability of stage~I and~II CRC patients from whole-slide H\&E stained histopathology images. More specifically, the key aim is to predict whether the patient is likely to survive at least five years (the currently used clinically driven salient follow-up time) after surgery.

\subsection{Data preparation \& pre-processing}
The analysis of WSIs is widely conceived as one of the most challenging tasks in the field of medical image analysis due to the following factors: (i) individual image size, (ii) usually extreme class imbalance, (iii) low total slide count (contextually speaking), (iv) scanning and preparation image artefacts, and (v) WSI salient information heterogeneity. We will shortly expand in greater detail but in summary, in order to address these we propose a sequence of steps which include chromatic normalization, patch extraction, data augmentation, and patch clustering.

\vspace{-0.5pt}\paragraph{Chromatic normalization}
Histopathological tissue sections or WSI are often examined individually by pathologists, who mainly focus on relative  colour and pattern differences within a single tissue section. It is rare to  compare directly different slides in order to make a diagnosis; each slide is examined to identify particular spatial or pattern characteristics. However, in the application of quantitative analysis and medical statistics for diagnosis and prognosis, different overall absolute colour value can lead to serious bias especially when the slide count is low. The variation in terms of colour distribution is ultimately the difference in the amount of light absorbed; Fig~\ref{f:normEx} shows examples of slides from different surgeries. As can be seen, the colour profile exhibits great inter-clinic variability. While it is true that the use of greyscale would address this problem, it also effects a loss of valuable pathological information.

We apply the Reinhard normalization \cite{reinhard2001color} which begins by converting the RGB representation into a perception-based colour space $l\alpha\beta$ with known phosphor chromaticity. This is done by a conversion from RGB to XYZ tristimulus values and then a conversion from XYZ space to LMS which is then followed by principal component analysis (effecting axes rotation).

The first conversion is based on the phosphors of monitor that is used to display an image. A device independent conversion is applied as an approximation that maps the white in the chromaticity diagram to white in RGB space:
{\small\begin{align}
  \left[
  \begin{array}{c}
       X\\
       Y\\
       Z
  \end{array}\right]=
  \left[
  \begin{array}{ccc}
    0.5141 & 0.3239 & 0.1604\\
    0.2651 & 0.6702 & 0.0641\\
    0.0241 & 0.1228 & 0.8444
  \end{array}\right]
  \left[
  \begin{array}{c}
       R\\
       G\\
       B
  \end{array}\right].
\end{align}}
Next, mapping to LMS space is performed:
{\small\begin{align}
  \left[
  \begin{array}{c}
       L\\
       M\\
       S
  \end{array}\right]=
  \left[
  \begin{array}{ccc}
    0.3897 & 0.6890 & −0.0787\\
   −0.2298 & 1.1834 &  0.0464\\
    0.0000 & 0.0000 &  1.0000
  \end{array}\right]
  \left[
  \begin{array}{c}
       X\\
       Y\\
       Z
  \end{array}\right].
\end{align}}
To eliminate distribution skew in the LMS space a logarithmic transform is applied and a pre-calculated maximal de-correlation matrix used to rotate the axes:
{\small\begin{align}
  \left[
  \begin{array}{c}
       l\\
       \alpha\\
       \beta
  \end{array}\right]=
  \left[
  \begin{array}{ccc}
    \frac{1}{\sqrt{3}} & 0 & 0\\
    0 & \frac{1}{\sqrt{6}} & 0\\
    0 & 0 & \frac{1}{\sqrt{2}}
  \end{array}\right]
  \left[
  \begin{array}{ccc}
    1 &  1 &  1\\
    1 &  1 & -2\\
    1 & -1 &  0\\
  \end{array}\right]
  \left[
  \begin{array}{c}
       L\\
       M\\
       S
  \end{array}\right].
\end{align}}
where the $l$ component is the achromatic value, the $\alpha$ and $\beta$ components represent chromatic yellow-blue and red-green channel values. Now the achromatic axis is orthogonal to the equiluminant plane. Finally, the distribution of values in each channel is normalized to be the standard normal.

\vspace{-0.5pt}\paragraph{Patch extraction}
Patches are extracted after the entire WSI has been down-sampled and normalized. The gigapixel resolution of the WSIs makes the existing approaches in the literature prohibitively computationally demanding. To make the model trainable, herein tiles of size $224 \times 224$ pixels are extracted from the 1/10 resolution image (n.b.\ $40\times$ magnification level was used in acquisition).

Recall that the patch extraction process in the pioneer work of \cite{hou2016patch} is still human labour intensive as it requires
patches with less than 30\% of tissue area or excessive blood content to be labelled and discarded manually. To construct an end-to-end pipeline, we instead approach the problem of relevant patch selection through the use of fully automatic clustering which does not assume or require application of human prior knowledge.

\vspace{-0.5pt}\paragraph{Data augmentation}
To prevent the model from over-fitting and alleviate problems caused by data imbalance, data augmentation is applied to the
training data. Herein we perform augmentation by means of applications of a restricted set of affine transformations, Gaussian blur, and all-channel multiplication.

Specifically, we apply geometric transformations by means of 90$^\circ$ rotations, and vertical and horizontal reflections, in random combinations, effecting an 8-fold increase in the data set size. Gaussian blur with the unitary standard deviation is applied in order to make the model generalize to blurred images since blur is often found in regions of WSIs as a result of locally poor focusing \cite{lopez2013automated}. 

\begin{figure*} 
  \centering 
  \includegraphics[width=.99\textwidth]{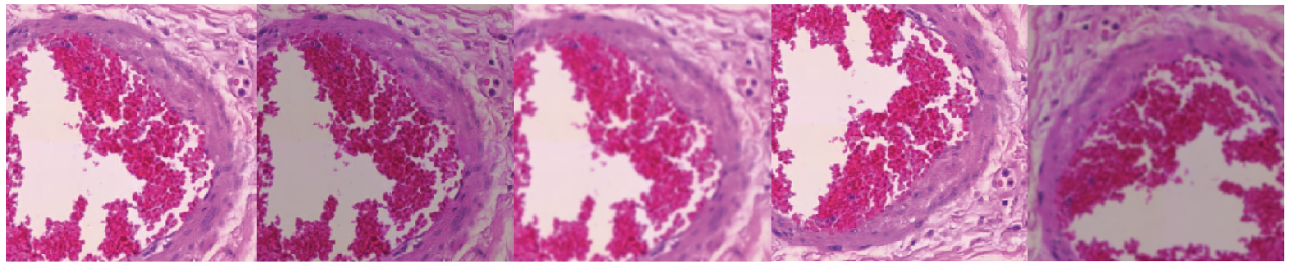}
  \caption{Training corpus augmentation by synthetically generated imagery. Shown is an original image (far left) and four examples of synthetic images generated from it (the remainder).}
  \label{f:augm}
\end{figure*}

\subsection{Patch clustering}
A major problem for patch-wise based classification approaches is that there is no ground truth label for each individual patch. In order to overcome this issue, we broadly consider a patch either to be (sufficiently) discriminative or not. This alone does not get one much further as it is very difficult to extract the discriminative subset of patches without expert knowledge and intensive human labour. Therefore, to obtain a collection of discriminative patches, we propose that an unsupervised learning method is used to cluster similar patches into several groups. In particular, we apply the $k$-means algorithm to group patches from a single WSI \cite{Aran2013c}. To increase the robustness of the result to the random initialization of parameters we perform multiple clusterings using different random starting parameters, and adopt the one associated with the lowest loss, thereby avoiding sub-optimal local minima. Lastly, CNN based classifiers are trained with patches from different clusters and used to determine which clusters are discriminative. In this work we adopt two clustering approaches, described next.

\paragraph{Information density clustering}
Information density is a simple but efficient way to group the extracted patches. In particular, since peripheral patches tend to contain large uniform areas, they are suited for compression by the DEFLATE algorithm used by the PNG image format. %[54]

The information ratio is defined here as the inverse of data compression ratio, $IR =\frac{1}{CR}=\frac{S_c}{S_u}$ where $S_u$ is the bit size of an uncompressed image. For a $224\times 224$ pixel RGB 8-bit image, $S_u$ is 150,528 bytes, and $S_c$ is the size of the corresponding losslessly compressed PNG file.

\begin{figure}
  \centering 
  \subfigure[]{
  \includegraphics[width=.47\textwidth]{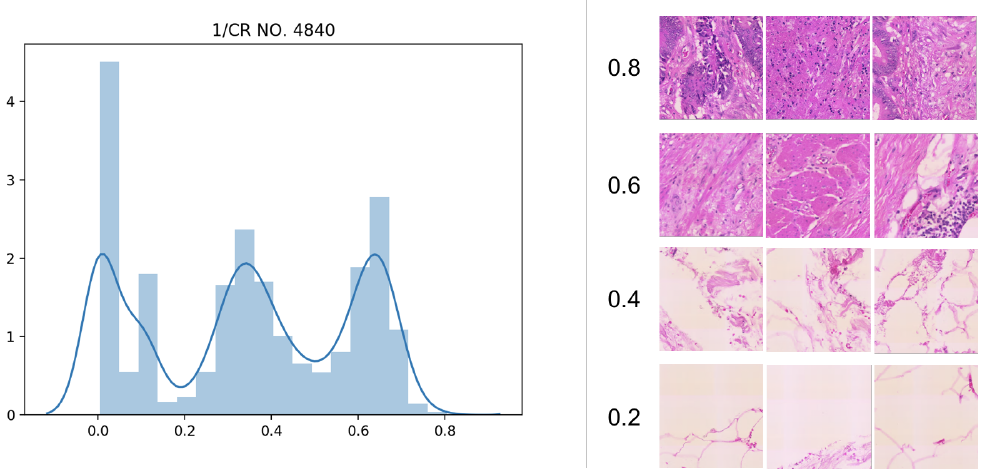}
  \label{f:clusteringCR}}
  \subfigure[]{
  \includegraphics[width=.47\textwidth]{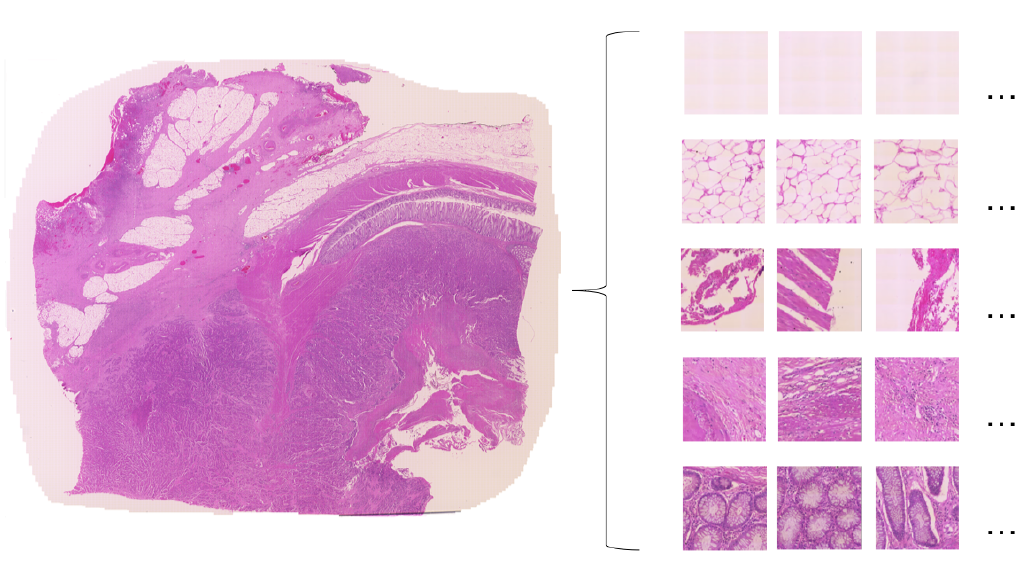}
  \label{f:tiles}}
  \caption{(a) Distribution of patch memberships (left) across different information ratio based clusters and the corresponding visual examples (right). (b) An example of a WSI (left) and sample patches (right) from the different phenotype based clusters inferred automatically. %from the slide.
  }
\end{figure}

An example is shown in Fig~\ref{f:clusteringCR}. Observe that most patches fall into one of the three clusters (${CR}^{-1} = 0.1, 0.4, 0.7$) as well as that patches with higher ${CR}^{-1}$ tend to contain more salient (cancer related) information than others. However, it is important to note that this does not necessarily imply that they are more pertinent for \emph{prognosis} i.e.\ our ultimate task. For example, the spatial arrangement of immune and cancer cells in peripheral regions around the tumour is known to be informative in this regard.

\begin{figure} 
  \centering 
  \includegraphics[width=.48\textwidth]{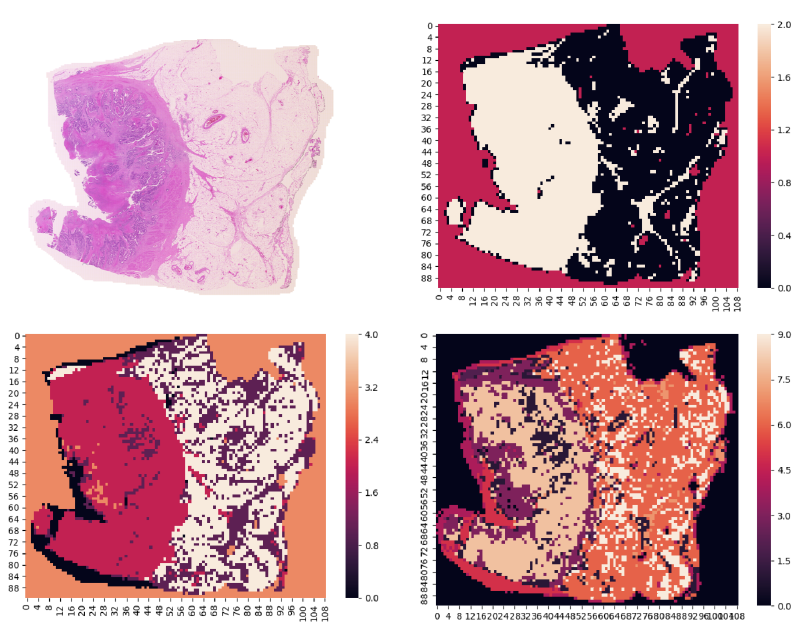}
  \includegraphics[width=.48\textwidth]{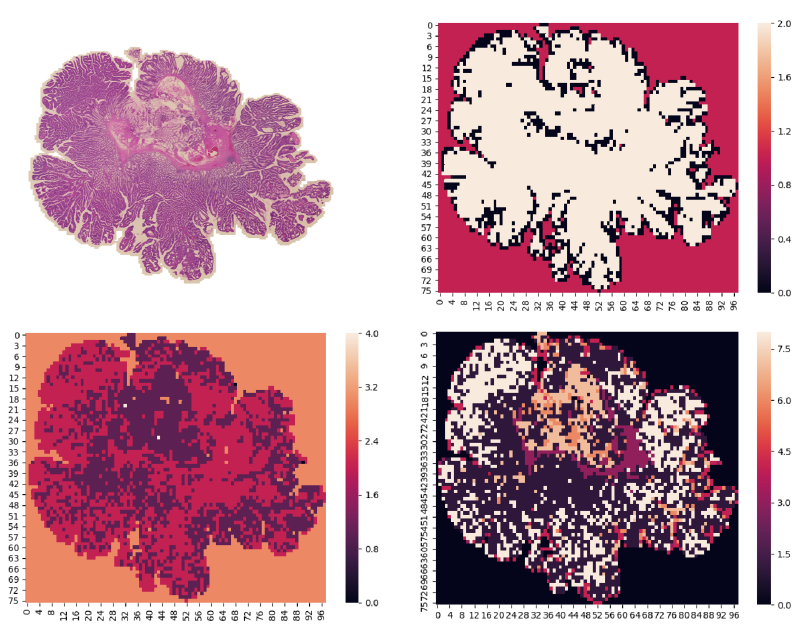}%\\
  \caption{Examples of clustering results. Each block of four images comprises the original WSI (top left), and the corresponding information density (top right) and phenotype based ($k=5$ and $k=10$ respectively bottom left and bottom right) clustering labels, colour coded.}
  \label{f:clusteringPhenotype}
\vspace{-0.8pt}\end{figure}

\paragraph{Phenotype clustering}
We also developed a new \emph{phenotype clustering} approach \cite{zhu2017wsisa}. The motivation behind phenotype clustering stems from the observation that the extracted patches exhibit significant heterogeneity; see Fig~\ref{f:tiles}. Because it is computationally expensive to perform clustering in the original 150,528 dimensional space ($224 \times 224 \times 3$), herein (instead of performing simple down-sampling) we used an ImageNet pre-trained CNN to generate phenotypes and then principal component analysis for dimensionality reduction. A summary of the process is shown in Table~\ref{t:clusteringPhenotype} and visual examples in Fig~\ref{f:clusteringPhenotype}.

\begin{SCtable}
  \centering
  \renewcommand{\arraystretch}{1.0}
  \caption{Summary of data flow and transformation at different stages of the proposed algorithm employing phenotype clustering based patch selection.}
  \vspace{5pt}
  \begin{tabular}{l|c}
    \Hline
    Operation & Input dimension\\
    \hline
    Pre-trained CNN &  $224 \times 224 \times 3$\\
    Global average pooling & $7 \times 7 \times 512$\\
    Dimension reduction (PCA) & $512$\\
    $k$-means clustering & $50$\\
    \Hline
  \end{tabular}
  \label{t:clusteringPhenotype}
\end{SCtable}

\vspace{-0.5pt}\subsection{Patch-wise CNN prognosis}
Our CNN design was inspired by the well-known VGG16 network though with some notable changes to its architecture and training, made to best suit the problem at hand. The main of these are: (i) size of the fully connected layer, (ii) learning optimizer, (iii) learning rate, and (iv) use of transfer learning.

We apply the CNN on the patch level and regard the ground truth label of the entire corresponding WSI as its label. The fact that not all patches contain discriminative features (for survival prediction) motivated our choice to train a network for each cluster separately and independently. Then the clusters which correspond to the networks that converge and have high validation accuracy are inferred to be discriminative. Networks of non-discriminative patches clusters may also converge but their high validation error still allows for the automatic inference of their non-discriminative nature.

\subsection{Aggregation of predictions}
The validation accuracy threshold of 65\% was chosen for the selection of discriminative clusters. After the patch levels predictions are made, these are aggregated into cluster level predictions. Since each WSI has different numbers of tiles from a given cluster, patch level predictions are represented by normalized histograms, thus effecting a homogeneous representation. A support vector machine (SVM) classifier \cite{BirkAranHump2017} is trained to learn the cluster level outcome.

\vspace{-0.5pt}\section{Experimental evaluation}\vspace{-0.3pt}

\subsection{Patient samples and ethics}
Whole slide images from patients operated on in NHS Lothian hospitals were used in the present work. Our data set comprises WSIs of tissue sections from each of the diagnostically residual and archived formalin fixed paraffin embedded tissue blocks, from CRC stage I and stage II patients who underwent surgical resection. This work was conducted in accordance with the declaration of Helsinki and no patient identifiable information was provided to the researchers. Ethical approval was obtained after review by the NHS Lothian NRS BioResource, REC approved Research Tissue Bank (REC approval ref: 15/ES/0094), granted by East of Scotland Research Ethics Service. Apart from the WSIs, each patient data sample is accompanied by a demographical description and follow-up information including age at surgery, date of death, and whether this patient dies of CRC (recall that in the present paper we are interested in outcome prognosis i.e.\ survivability prediction). The original slides were stained using haematoxylin and eosin at the time of treatment, and were scanned using a ZEISS Axio scan Z1 (Zeiss, Oberkochen, DE) whole slide scanner with a $40\times$ objective. The scale of a single pixel represents $0.111\mu m \times 0.111\mu m$ of the actual size. The digital camera used was a Hitachi HVF2025SCL with an exposure time of 200$\mu s$. As our focus is on prognostic classification, all the images in the corpus correspond to positively diagnosed CRC cases.

The entire data set contains 75 WSIs representing one slide from a single patient. The smallest image is 6GB and the largest one is 18GB, with an average size of the WSIs being 8GB, which is approximately 300,000 pixels by 200,000 pixels. The bit depth is 24 with 3 channels. The original data was in the commonly used CZI format and included an identity label for each slide.

\vspace{-0.3pt}\subsection{Results and discussion}\vspace{-0.2pt}
We started our analysis by examining the overall prediction results and, in particular, the effect that different clustering and aggregation techniques, and their parameters have. For comparison, in addition to our SVM based aggregation described in the previous section, we also present results for majority voting based aggregation \cite{AranCipo2004a}. A summary is provided in Table~\ref{t:results}.

\begin{SCtable}
  \centering
  %\small
  \renewcommand{\arraystretch}{1.0}
  \caption{Summary of results; prefixes ID and Ph refer to respectively information density and phenotype based clustering, followed by the corresponding number of clusters.}
  \vspace{5pt}
  \begin{tabular}{l||cc|cc}
    \Hline
       & \multicolumn{2}{c|}{Patch level} & \multicolumn{2}{c}{Cluster level}\\
                & Accuracy    & F1  & Accuracy    & F1\\ 
    \hline
      ID3-CNN-Vote  & 0.60 & 0.67 & 0.50 & 0.67\\
      ID3-CNN-SVM   & 0.60 & 0.67 & 1.00 & 1.00\\
      Ph5-CNN-Vote  & 0.68 & 0.67 & 1.00 & 1.00\\
      Ph5-CNN-SVM   & 0.68 & 0.67 & 1.00 & 1.00\\
      Ph10-CNN-Vote & 0.70 & 0.81 & 0.50 & 0.67\\
      Ph10-CNN-SVM  & 0.70 & 0.81 & 0.50 & 0.67\\
    \Hline
  \end{tabular}
  \label{t:results}
\end{SCtable}

There are several important observations that are readily apparent from the table. Firstly, all of the approaches -- that is, different combinations of clustering and decision fusion techniques -- performed very well indeed already on the level of individual patches. On the patch level phenotype based clustering with larger $k$ performed best. Interestingly, at this stage the manner of decision fusion (majority vote vs.\ SVM based) made no difference in the context of any of the different algorithms. 

Further insight can be gained by looking at cluster level performance, especially when interpreted in the context of the aforementioned results. Here we do observe some advantage of SVM based decision fusion, albeit only when information density based clustering is employed. Another noteworthy observation is that unlike in other cases, cluster level prediction is worse when phenotype based clustering with larger $k$ is used. Our hypothesis, which requires further experiments for validation, is that this is not a `true' trend but rather a stochastic anomaly which emerges from the need of more data for large $k$. Lastly, and most importantly, moving from patch to cluster level prediction using SVM based fusion dramatically improves algorithms with both information density and small $k$ phenotype based clustering, and results in perfect performance. This observation also supports our hypothesis as regards the anomaly noticed for large $k$ phenotype based clustering.

Finally, we sought additional insight and examined the $k=5$ clusters for the best performing method (Ph5-CNN-SVM). What we found was that the semantics of the five clusters were very easy to interpret and can be summarized as containing the following patch types --
Cluster 0: outside of the tissue micro-section,
Cluster 1: containing blood cells,
Cluster 2: containing cancerous and immune cells,
Cluster 3: void, and
Cluster 4: containing fat cells.
Thus, we can conclude that our method not only performs extremely well in terms of the ultimate goal of survival prognosis but also that it does so by learning clinically meaningful problem structure. Ample previous work testifies to the importance of interpretability in the adoption of novel machine learning assistive tools by medical professionals.

\vspace{-0.5pt}\section{Summary and conclusions}\vspace{-0.3pt}
This paper is the first work to address one of the most challenging problems in the emerging sphere of digital pathology -- that of using images not previously annotated by a pathologist to develop algorithms that can be applied automatically to generate diagnostic and prognostic information from WSIs. Almost all current applications of CNN require careful annotation of tissue images by a qualified pathologist, and this is a rate limiting step. The novel algorithm we introduced addresses the overwhelming amount of data by automatic, unsupervised discriminative patch selection and the convergence performance of cluster level trained convolutional neural networks, and the inference of prognosis on the level of individual discriminative clusters followed by decision fusion using support vector machines. On a real-world corpus our phenotype based clustering employed in conjunction with the aforementioned techniques achieved perfect performance both in terms of overall accuracy and F1 score.

\section*{Acknowledgements}
The authors gratefully acknowledge the support of NVIDIA Corporation for their donation of the Titan Xp GPU used for this research.

\tiny

\label{sect:bib}
\bibliographystyle{plain}
\bibliography{./2019_BICOB2_ab}

\end{document}